\documentstyle[psfig,11pt]{article}

\begin{document}
\begin{small}
\noindent{\it Radiophysics and Quantum Electronics, Vol. 41, No. 8, 1998, 669--673}
\end{small}

\vskip 1cm
\noindent{\bf\Large A DIFFERENTIAL METHOD OF MAXIMUM ENTROPY}

\bigskip

\begin{large}
\begin{quote}

{\bf A.T.Bajkova}

\bigskip

{\it
We consider a differential method of maximum entropy that is based on
the linearity of Fourier transform and involves reconstruction of images
from the differences of the visibility function. The efficiency of the
method is demonstrated with respect to the recovery of source images
with bright components against the background of a sufficiently weak
extended base. The simulation results are given along with the maps of
an extragalactic radio source 0059+581, which were obtained using the
standard and differential methods of maximum entropy for three
observation dates and show that the principle of differential mapping
allows us to increase considerably the dynamic interval of images.}
\end{quote}

\vskip 1cm

\noindent{\bf 1.~THE PRINCIPLE OF DIFFERENTIAL MAPPING}

\bigskip

The idea of differential mapping is not new. At present, it is realized
most comprehensively in the program package "DifMap" [1] at the
California Institute of Technology, where the "cleaning" procedure
(CLEAN) is used as deconvolution operation.

The method of differential mapping is based on the fundamental
property of the Fourier transformation. According to this method, the bright
components of the source, which were reconstructed at the first stage,
are subtracted from the initial visibility function, the subsequent
reconstruction is performed using the residual visibility function, and
the reconstruction results are added at the final stage.

In the case of CLEAN, such a mapping method mainly influences the
convergence rate. However, if the maximum entropy method (MEM), which has
pronounced nonlinear
properties, is used as the deconvolution operation, the principle of
differential
mapping can improve the reconstruction quality, particularly when the
source has bright compact components against the background of a
sufficiently weak extended base.
An example of such improvement due to subtraction of the bright
component from the visibility function is shown in [2].

What is the cause of the reconstruction quality improvement? The fact is
that after the subtraction of bright components, which were
reconstructed at the first stage using the MEM, from the initial
visibility function we obtain the residual visibility function, in which
the proportion of the weak extended component becomes larger. (The
subtraction result is called the first-order residual visibility
function.) Therefore,
we artificially decrease the dynamic interval of the map, which
corresponds to the residual visibility function, and, thus simplify the
image reconstruction at the second stage.

If the bright components of the image reconstructed at the second stage
are subtracted from the first-order residual visibility function, we
obtain a second-order residual visibility function, to which a map with
a still lower dynamic interval corresponds. Using this procedure, we
obtain residual visibility functions of an increasingly high order.

To obtain the desired source map, the map reconstructed at the last
stage must be summed up with all the components subtracted from the
visibility function at the previous stages.

Formally, the algorithm of two-stage differential mapping is formulated
as

$$
F_{vis}^{(2)} = F_{vis}^{(0)} - F_{vis}^{(1)},
$$
where  $F_{vis}^{(0)}$ is the initial visibility function of the desired
brightness distribution over the source $I(x)$; $F_{vis}^{(1)}$ is the
visibility function corresponding to the bright component $I^{(1)}(x)$
reconstructed at the first stage; $F_{vis}^{(2)}$ is the residual
visibility function corresponding to the source map $I^{(2)}(x)$
reconstructed at the second stage.

The resulting map has the form

$$
I(x)=I^{(1)}(x)+I^{(2)}(x).
$$

\bigskip

\noindent{\bf 2.~SIMULATION RESULTS}

\bigskip

In Figs.~1 and 2 we show the results of simulation of the principle of
differential mapping using the generalized maximum entropy method (GMEM)
[3] for real images with both positive and negative values (the advantages of the GMEM compared with the MEM are discussed
below in Section 3).

The experiment shown in Fig.~1 was carried out for three model sources
in the form of one or several bright point components against the
background of a weaker extended base that has a Gaussian form. In the
figures we give the following source names: "Model-1", "Model-2", and
"Model-3". For "Model-1" the base amplitude is 20\% of the amplitude of
the bright point-like component, while for "Model-2" it is only 2\%.
"Model-3", which has four bright components against the background of a
20\% Gaussian, is more complicated. To form the visibility function, we
used the symmetric coating of the $UV$ surface with almost uniform
filling over the entire aperture, which included 98 points. To study the
reconstruction improvement due to use of the differential mapping
principle in its pure form, no noise was added to the visibility
function.

\begin{figure}
\centerline{
\psfig{figure=model1.ps,angle=-90,width=100mm}}
\centerline{Fig.1.}
\end{figure}

\begin{figure}
\centerline{
\psfig{figure=model2.ps,angle=-90,width=100mm}}
\centerline{Fig.2.}
\end{figure}

In Figs.~1 and 2, for each vertically located source, we show the maps
that appear horizontally in the following order (from left to right):
the map of the model source, a "noisy" image resulting from the inverse
Fourier transform of the initial visibility function, an image
reconstructed at the first stage of differential mapping using the GMEM,
and the resulting map of differential mapping. (It must be noted that
the bottom level of the contour line corresponds to 1\% of the peak
value of the flux in all the above images.)

Obviously, the maps obtained at the first stage can be treated as the
limiting performance of the GMEM applied to the initial visibility
function. The subtraction of the bright components reconstructed at the
first stage from the visibility function allows us to improve
significantly the reconstruction quality at the second stage. In Table 1
we show the quantitative properties of the image reconstruction quality
from Fig.~1 after the first and second stages of the differential
mapping.

As is obvious from the table, compared with the standard GMEM, the use
of the differential GMEM allows us to increase the signal-to-noise ratio
(SNR) of the output image by a factor of 3 to 5, such that the maximum
effect is achieved for the source "Model-2", which has maximum dynamic
interval.

In Fig.~2 we show the results of reconstruction of images of the sources
"Model-1" and "Model-2" for the case of another filling of the $UV$
plane, which includes 98 points distributed uniformly inside the central
region. The above region is approximately a quarter of the aperture.
This example of mapping is more complicated compared with the previous
case because of the total absence of the high spatial harmonics of the
source in the visibility function. As is obvious from the figure, in
this case the differential GMEM also gives higher-quality results than
the standard GMEM.

\medskip
\centerline{Table~1}
\medskip
\centerline{Characteristics of reconstructed images}
\begin{center}
\begin{large}\begin{tabular}{lllll}
\hline
\noalign{\vskip 1mm}
        & Maximum    &   SNR      & Maximum   & SNR  \\
Source  & error      &            & error      &       \\
        & at the 1st stage & at the 1st stage & at the 2nd stage & at the 2nd stage\\
\noalign{\vskip 1mm}
\hline
\noalign{\vskip 1mm}
 Model-1 & 0.4701  &   3.84 & 0.0572  &  15.86 \\
 Model-2 & 0.0467  &  11.43 & 0.0088  &  52.00 \\
 Model-3 & 0.1890  &   4.10 & 0.0594  &  13.20 \\
\noalign{\vskip 1mm}
\hline
\end{tabular}\end{large}\end{center}

\bigskip

\noindent{\bf 3.~MAPPING FOR THE RADIO SOURCE 0059+581}

\bigskip

To demonstrate the potentialities of the differential generalized method
of maximum entropy used for the real data processing, we show the results of
mapping the extragalactic radio source 0059+581, which is often used as
a reference source in the VLBI geodetic programs [4]. This source is of
astrophysical interest because it shows fast variation of both the total
flux and the structure. The maps that are constructed using the GMEM for
a number of dates in the interval from June 1994 to December 1995 are
shown in Fig.~3. Variation of the total flux in the same interval is
presented in Fig.~4 in [5].

In the time interval from the middle to the end of 1994 we observed an
almost linear decrease in the total flux from $\sim$ 4 Ja to 1.5 to 2 Ja.
As is obvious from Fig.3, we failed to obtain a map with dynamic range
sufficient for detecting an extended element from the data of June 26,
1994 when the flux was maximal over the interval in question. Using the
standard GMEM, we detected extended elements only starting from October
4, 1994 when their share with respect to the total flux became sufficient
to be detected by the selected reconstruction method. In Fig.~4, for
comparison, we show maps obtained using both the standard and
differential GMEMs for the three dates on which the total flux was
sufficiently large and the flux share corresponding to extended elements
was insignificant. As is obvious from the figure, the use of
differential mapping allowed us to increase the dynamic range of the
maps such that extended structures became very distinguished. It must be
noted that the minimal level of the contour line corresponds to 1\% of
the peak value of the flux everywhere in the images. The parameters of
the maps obtained using the differential GMEM are given below in Table~2.

\begin{figure}
\centerline{
\psfig{figure=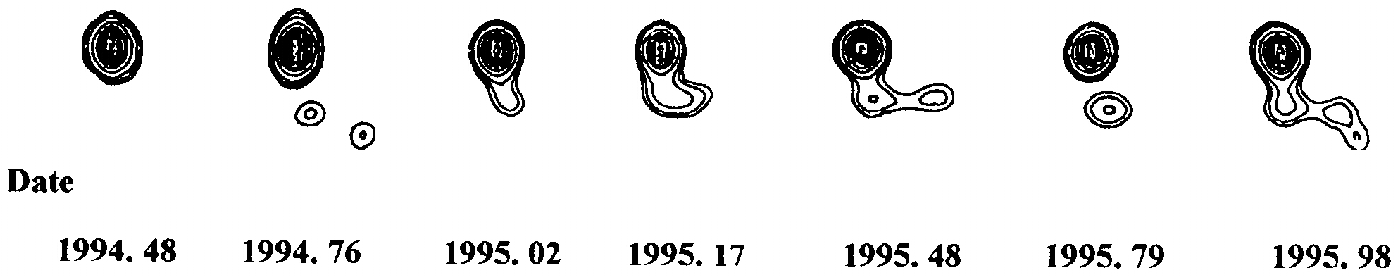,width=150mm}}
\centerline{Fig.3.}
\end{figure}

\begin{figure}
\centerline{
\psfig{figure=0059.ps,angle=-90,width=100mm}}
\centerline{Fig.4.}
\end{figure}

\medskip
\centerline{Table~2}
\medskip
\centerline{Map parameters for the source 0059+581}
\begin{center}
\begin{tabular}{lll}
\hline
\noalign{\vskip 1mm}
Date& Total   & Peak        \\
    & Flux(Ja)& flux(Ja) \\
\noalign{\vskip 1mm}
\hline
\noalign{\vskip 1mm}
 25.06.94 & 4.23 & 2.40 \\
 25.08.94 & 3.40 & 1.76 \\
 04.10.94 & 3.28 & 1.86 \\
\noalign{\vskip 1mm}
\hline
\end{tabular}\end{center}

The main criterion for choosing the method of maximum entropy (either
classical or generalized, the MEM or the GMEM) is the type of the
desired distribution. If the desired image is real bipolar (with
positive and negative values) or complex
with both real and imaginary parts, only the GMEM [6] can be used.

If it is known a priori that the image is real and nonnegative, the
classical MEM  can be used. However, one should keep in mind that in
case of large errors in the data, the GMEM allows us to obtain [7]
higher-quality maps with a much lower level of nonlinear distortions, as
compared with the MEM. At present, the researchers of the IAA RAS carry
out research in making maps of the extragalactic sources using the
observation data from the astrometric and geodetic VLBI programs. These
data were not obtained directly for astrophysical mapping, and, as a
rule, have poor calibration. Therefore, to obtain satisfactory maps, the
GMEM is recommended instead of the MEM.

Another advantage of the GMEM originates from the fact that an image
with negative values can correspond to the residual visibility function
obtained after the subtraction of the bright component that is
reconstructed at the first stage of the differential algorithm from the initial
visibility function. This is possible if the bright component was
reconstructed at the first stage with an overestimated amplitude, which
is quite typical of any nonlinear method. To rule out the unwanted image
distortions when a nonnegative solution is sought using the data
assuming the presence of negative components, we should use the
generalized method of maximum entropy for real bipolar images rather
than the classical method.

\bigskip

\noindent{\bf 4.~CONCLUSIONS}

\bigskip

Compared with the traditional methods, the nonlinear methods of
differential mapping using the maximum entropy method as the
deconvolution operation allow us to improve substantially the
reconstruction quality of source images containing bright point-like
components against the background of weak base. The maps obtained using
the differential method of maximum entropy are characterized by a higher
dynamic range. To eliminate possible nonlinear distortions in the case
of differential mapping, the generalized method of maximum entropy is
preferred to the classical method.

\bigskip

This work was supported by the Russian Foundation for Basic Research,
grant No 96-02-19177 and the Ministry of Science's program "Astronomy.
Fundamental Space Research", grant No 2.1.1.3.

\bigskip

\end{large}

\bigskip
\noindent{\bf REFERENCES}
\bigskip

\noindent{1.~G.Taylor, {\it The Difmap Cookbook}, California Institute of Technology,
Pasadena (1994).}

\noindent{2.~T.Cornwell, "Very Long Baseline Interferometry and
the VLBA", in: {\it ASP Conference Series} (J.A. Zensus, P.J. Diamond,
and P.J. Napier, eds) vol.82, 227 (1995).}

\noindent{3.~A.T.Bajkova, {\it Commun. IAA RAS}, No.~58 (1993).}

\noindent{4.~A.T.Bajkova et al, {\it Trans. IAA RAS}, No.~1, 22 (1997).}

\noindent{5.~T.B.Pyatunina, A.T.Bajkova et al, {\it Trans. IAA RAS}, No.~1,
64 (1997).}

\noindent{6.~A.T.Bajkova, {\it Izv. Vyssh. Uchebn. Zaved., Radiofiz.},
{\bf 34}, No. 8, 919 (1991).}

\noindent{7.~A.T.Bajkova, {\it Izv. Vyssh. Uchebn. Zaved., Radiofiz.}, {\bf
38}, No. 12, 1267 (1995).}
\end{document}